\documentclass[conference]{IEEEtran}
\IEEEoverridecommandlockouts

\usepackage{url}
\usepackage{cite}
\usepackage{amsmath,amssymb,amsfonts}
\usepackage{algorithmic}
\usepackage{graphicx}
\usepackage{textcomp}
\usepackage{xcolor}
\def\BibTeX{{\rm B\kern-.05em{\sc i\kern-.025em b}\kern-.08em
    T\kern-.1667em\lower.7ex\hbox{E}\kern-.125emX}}
\begin{document}

\title{Probabilistic Symbol-Level Precoding based Affine Frequency Division Multiplexing Transmission}

\author{
	\IEEEauthorblockN{Shuntian Tang\IEEEauthorrefmark{1}, 
		Xinyi Wang\IEEEauthorrefmark{1}, 
		Shuangyang Li\IEEEauthorrefmark{2}, 
		Tianqi Mao\IEEEauthorrefmark{3}, 
		Zilong Liu\IEEEauthorrefmark{4}, and 
		Zesong Fei\IEEEauthorrefmark{1}}
	\IEEEauthorblockA{\IEEEauthorrefmark{1}School of Information and Electronics, Beijing Institute of Technology, China} 
	\IEEEauthorblockA{\IEEEauthorrefmark{2}Faculty of Electrical Engineering and Computer Science, Technical University of Berlin, Germany} 
	\IEEEauthorblockA{\IEEEauthorrefmark{3}the State Key Laboratory of Environment Characteristics and Effects for Near-space, Beijing Institute of Technology, China} 
	\IEEEauthorblockA{\IEEEauthorrefmark{4}School of Computer Science and Electronics Engineering, University of Essex, U.K} 
	
	\IEEEauthorblockA{
		Email: bit\_tangshuntian@bit.edu.cn, wangxinyi@bit.edu.cn, shuangyang.li@tu-berlin.de, \\ maotq@bit.edu.cn, zilong.liu@essex.ac.uk, and feizesong@bit.edu.cn}
}

\maketitle

\begin{abstract}
Affine frequency division multiplexing (AFDM) has recently gained significant attention due to its robustness against time-frequency doubly selective channel fading. However, the high computational complexity at the receiver poses a critical challenge for practical deployment. To overcome this issue, we propose a probabilistic symbol-level precoding (SLP)-based AFDM transmission framework, in which the processing burden in downlink transmission is shifted from the user to the base station (BS), enabling direct symbol detection without channel estimation or equalization at the receiver.
In the proposed framework, the BS exploits the uplink channel state information (CSI) to design the downlink transmit waveform based on uplink–downlink channel reciprocity. In particular, we innovatively introduce a probabilistic SLP technology by explicitly characterizing the likelihood of symbol detection errors under noise perturbations. Specifically, the transmitted symbols are optimized to minimize the likelihood that the received symbols fall into erroneous decision regions, where the resulting error-probability minimization problem is subsequently approximated as a second order cone programming (SOCP) problem by exploiting the monotonicity of the objective function. Simulation results show that the proposed probabilistic SLP-based scheme achieves performance comparable to that of conventional AFDM receivers, whilst enjoying significant reduction of computational complexity at the receiver end. These results demonstrate the effectiveness and practical potential of the proposed approach.
\end{abstract}


\section{Introduction}
The sixth-generation (6G) wireless communication systems are envisioned to support diverse services and applications, including reliable data transmission over high-mobility channels and/or in high frequency bands~\cite{Saad2020NET}. The legacy orthogonal frequency division multiplexing (OFDM) suffers from severe performance degradation in the presence of strong Doppler shifts, carrier frequency offset, and phase noise due to loss of subcarrier orthogonality, making it unsuitable for 6G high-mobility scenarios, \textcolor{black}{like low Earth orbit (LEO) satellite communication~\cite{Kodheli2021COMST}, and intelligent high-speed railway (HSR) communication~\cite{Zhou2021TWC}.}

As a promising alternative, affine frequency division multiplexing (AFDM) has recently attracted considerable attention due to its inherent robustness against strong Doppler shifts~\cite{Bemani2023TWC, rou2026afdmevolvingofdm6g} and various hardware impairments~\cite{sui2026mimoafdmoutperformsmimoofdmface}. By leveraging the inverse discrete affine Fourier transform (IDAFT) with adjustable parameters, AFDM is capable of flexibly shaping the signal representation in the affine frequency (AF) domain to enable effective separation of multipath components in doubly dispersive channels. This property allows AFDM to achieve full diversity gain and maintain reliable performance even under severe time and frequency selectivity. Building upon these advantages, significant research attention has been devoted to enhancing AFDM from different perspectives. For multiple-input multiple-output (MIMO) aided AFDM, \cite{Luo2025TWC} proposed a joint data detection and decoding framework based on sparse graph theory. In~\cite{zhang2025nonorthogonalafdmpromisingspectrumefficient}, non-orthogonal AFDM was studied to improve the spectrum efficiency, while AFDM based index modulation was investigated to further enhance spectral efficiency and reliability in~\cite{Liu2025TWC} and~\cite{Liu2026WCL}. In addition, AFDM has demonstrated strong potential in emerging applications such as integrated sensing and communication (ISAC) as reported in~\cite{sui2025multi},~\cite{Zhang2026JSAC}, and~\cite{yin2025arxivAffine}. These studies collectively highlight the versatility and effectiveness of AFDM in addressing the challenges of next-generation wireless systems.

Although the aforementioned works have demonstrated the superior performance of AFDM, it is noted that conventional AFDM receivers typically incur high computational complexity. As a result, such designs are more  feasible at the base station (BS) in the uplink, but may be impractical for user equipment in the downlink due to their limited computational capabilities. Following this philosophy, a promising approach is to exploit the available \textcolor{black}{uplink channel state information (CSI) under the time division duplexing (TDD) operation mode to perform transmit signal shaping at the BS prior to downlink transmission}. \textcolor{black}{For example, the delay-Doppler alignment modulation proposed in~\cite{Lu2024TWC} enables the concurrent arrival of multipath components, but generally requires multiple antennas. Similarly, symbol-level precoding (SLP) techniques in~\cite{Li2018TWC, Alodeh2017TWC, Wang2024WCL}, which are based on geometric constellation shaping, have demonstrated significant performance gains and low-complexity reception. However, these approaches inherently rely on multi-antenna processing and thus are not directly applicable to single-input single-output (SISO) systems, which limits their practicality in simpler transceiver architectures.}

To address the aforementioned issues, we propose a probabilistic SLP-based AFDM transmission framework in the AF domain, \textcolor{black}{which is applicable to SISO systems}. Our key innovation hereby is to shift the signal processing burden from the receiver to the transmitter, thereby enabling low-complexity symbol detection at the downlink receiver end. Specifically, the corresponding waveform design is formulated as a probabilistic optimization problem, where the transmit symbols are optimized based on the available CSI at the transmitter to maximize the probability that the received symbols fall within their correct decision regions. By exploiting \textcolor{black}{the monotonicity of the objective function, the original problem is further approximated as a tractable convex optimization problem that can be solved via second-order cone programming (SOCP).}
The numerical results validate the effectiveness of the proposed user-equalization-free AFDM transmission framework. Our results show that it achieves performance comparable to the traditional minimum mean square error (MMSE) method but with much reduced complexity under various CSI conditions, confirming its robustness and practical feasibility. 

\section{Basic Principles of AFDM}
\label{Preliminaries}
We briefly review the basic principles of AFDM following the framework in \cite{Bemani2023TWC}. Consider an $N \times 1$ complex symbol vector $\mathbf{x}=[x_0,x_1,\cdots,x_{N-1}]^T$, whose elements are drawn from a M-ary constellation alphabet. By applying the $N$-point IDAFT, the symbol vector is transformed from the AF domain to the time domain. The resulting time domain samples are expressed as
\begin{equation}
	s_n = \frac{1}{\sqrt{N}}\sum_{m=0}^{N-1}{x_m e^{j2\pi\left(c_1n^2+\frac{1}{N}mn+c_2m^2\right)}}, n = 0,1,\cdots,N-1,
\end{equation}
where $c_1$ and $c_2$ are adjustable parameters that enable AFDM to achieve full diversity over doubly selective fading channels.

To maintain periodicity of the signal in the AF domain, a chirp-periodic prefix (CPP) of length $L_c$ is appended to the transmitted signal. The CPP samples are defined as
\begin{equation}
	s_n = s_{N+n}e^{j2\pi c_1\left(N^2+2Nn\right)}, n = -L_c,\cdots,-1,
\end{equation}
where $L_c$ is chosen to be larger than the maximum channel delay spread (measured in samples). It is worth noting that the CPP reduces to a conventional cyclic prefix (CP) when $2Nc_1$ is an integer and $N$ is even.

Assuming a wireless channel consisting of $P$ propagation paths, the received time domain signal can be written as
\begin{equation}
	r_n = \sum_{i=1}^{P}{h_i s_{n-l_i}e^{j2\pi f_i n}} + w_n, n = 0,1,\cdots,N-1,
\end{equation}
where $h_i$, $l_i$, and $f_i$ denote the complex channel gain, the discrete propagation \textcolor{black}{integer} delay, and the Doppler shift of the $i$-th path, respectively. \textcolor{black}{Notably, fractional delays are ignored as the high sampling period allows path delays to be well approximated by their nearest sampling instants~\cite{Raviteja2018TWC, Liu2025JSAC}.} The term $w_n \sim \mathcal{CN}(0,\sigma^2)$ represents additive white Gaussian noise (AWGN).

At the receiver, the $N$-point discrete affine Fourier transform (DAFT) is applied to recover the signal in the AF domain. Denoting the resulting vector as $\mathbf{y}=[y_0,y_1,\cdots,y_{N-1}]^T$, the received AF-domain samples are given by
\begin{align}
	y_m = \frac{1}{\sqrt{N}}&\sum_{n=0}^{N-1}{r_n e^{-j2\pi\left(c_2m^2+\frac{1}{N}mn+c_1n^2\right)}}, \notag \\
	& \qquad\qquad \qquad m = 0,1,\cdots,N-1,
\end{align}

After removing the CPP, the AF-domain input–output relationship can be compactly written in matrix form as
\begin{equation} \label{IO-relation}
	\mathbf{y} =\sum_{i=1}^{P}h_i\mathbf{H}_i\mathbf{x}+\widetilde{\mathbf{w}}= \mathbf{H}_{\text{eff}}\mathbf{x}+\widetilde{\mathbf{w}},
\end{equation}
where $\widetilde{\mathbf{w}}\sim\mathcal{CN}(0,\sigma^2\mathbf{I}_N)$ has the same statistics as $\mathbf{w}$. The matrix $\mathbf{H}_i \in \mathbb{C}^{N\times N}$ represents the channel contribution of the $i$-th path and is given by

\begin{equation}
	\mathbf{H}_i=\mathbf{\Lambda}_{c_2}\mathbf{F}\mathbf{\Lambda}_{c_1}\mathbf{\Gamma}_i\mathbf{\Delta}_{f_i}\mathbf{\Pi}^{l_i}\mathbf{\Lambda}_{c_1}^H\mathbf{F}^H\mathbf{\Lambda}_{c_2}^H,
\end{equation}
where $\mathbf{F}$ denotes the $N$-point discrete Fourier transform (DFT) matrix, $\mathbf{\Lambda}_{c}=\text{diag}([1,e^{-j2\pi c},\cdots,e^{-j2\pi c(N-1)^2}])$, $\mathbf{\Delta}_{f_i}$ is the Doppler shift matrix, and $\mathbf{\Pi}$ denotes the cyclic permutation matrix. The diagonal matrix $\mathbf{\Gamma}_i$ is defined as
\begin{equation}
	\mathbf{\Gamma}_{i}(i,i)= \begin{cases}e^{-j 2 \pi c_1\left(N^2-2 N\left(l_i-n\right)\right)}, & n<l_i, \\ 1, & n \geq l_i.\end{cases}
\end{equation}
When $2Nc_1$ is an integer and $N$ is even, $\mathbf{\Gamma}_i$ simplifies to the identity matrix $\mathbf{I}_N$.

For the case of integer normalized Doppler shifts, the channel matrix $\mathbf{H}_i$ exhibits a sparse structure and its elements can be expressed as
\begin{equation} \label{channel_struc}
	\mathbf{H}_i(p, q)= \begin{cases}e^{j \frac{2 \pi}{N}\left(N c_1 l_i^2-q l_i+N c_2\left(q^2-p^2\right)\right)} & q=\left(p+\operatorname{loc}_i\right)_N, \\ 0 & \text { otherwise },\end{cases}
\end{equation}
where $\text{loc}_i = \alpha_i+2Nc_1l_i$ with $\alpha_i=Nf_i \in [-\alpha_{\text{max}}, \alpha_{\text{max}}]$ and $l_i \in [0, l_{\text{max}}]$. The parameters $\alpha_{\text{max}}$ and $l_{\text{max}}$ are defined as the maximum normalized Doppler shift and maximum normalized delay, respectively. As shown in \cite{Bemani2023TWC}, selecting $c_1 \geq \frac{2\alpha_{\text{max}}+1}{2N}$ ensures that each row and column of $\mathbf{H}_{\text{eff}}$ contains exactly $P$ nonzero elements.

For a more general scenario with fractional Doppler shifts, the elements of $\mathbf{H}_i$ can be written as
\begin{align}\label{gen_channel_struc}
	\mathbf{H}_i(p, q)&= \frac{1}{N}e^{j\frac{2\pi}{N}\left(Nc_1l_i^2-ql_i+Nc_2\left(q^2-p^2\right)\right)} \notag \\
	&\times \sum_{n=0}^{N-1}e^{-j\frac{2\pi}{N}\left(\left(p-q+v_i+2Nc_1l_i\right)n\right)},
\end{align}
where $v_i=Nf_i=\alpha_i+\nu_i$ denotes the normalized Doppler shift, with $\alpha_i$ and $\nu_i\in[-0.5,0.5]$ representing its integer and fractional components, respectively.
\section{System Model}
\label{System Model}

\textcolor{black}{In this paper, we propose an AFDM transmission framework that is free of channel estimation and equalization at the user side, while} major of signal processing tasks are handled at the BS. The overall procedure consists of two stages: an \textbf{uplink channel acquisition stage} and a \textbf{downlink precoding stage}.


During the uplink stage, the BS first acquires CSI from user transmissions. We consider a frame-based AFDM transmission structure and assume that the duration of each frame is sufficiently short compared with the relevant channel coherent time. Under this assumption, the sparsity pattern of the effective channel matrix $\mathbf{H}_{\text{eff}}$ remains unchanged within a frame.
Let $\mathbf{X}\in\mathbb{C}^{N\times L}$ denote an AFDM transmission frame containing $L$ symbols. The frame is constructed by placing a pilot symbol at the beginning, followed by $L-1$ data symbols, i.e.,
\begin{equation}
	\mathbf{X}= [\mathbf{x}_\text{p}, \mathbf{x}_{\text{d},1},\cdots,\mathbf{x}_{\text{d},L-1}].
\end{equation}
where $\mathbf{x}_\text{p}\in\mathbb{C}^{N\times1}$ represents the pilot signal and $\mathbf{x}_{\text{d},l}\in\mathbb{C}^{N\times1}$ denotes the $l$-th data symbol. The pilot symbol enables the BS to obtain an estimate of the effective channel matrix \textcolor{black}{using, e.g., the estimator proposed in~\cite{Tang2025ICCC}}, denoted by $\tilde{\mathbf{H}}_{\text{eff}}$. Using this estimated CSI, the BS subsequently performs equalization to recover the uplink data transmitted by the user.

After obtaining the CSI, the system proceeds to downlink transmission. Leveraging channel reciprocity between the uplink and downlink links~\cite{Ran2011ICCTA}, the BS utilizes the estimated channel matrix $\tilde{\mathbf{H}}_{\text{eff}}$ to design the downlink transmit waveform. \textcolor{black}{Considering the fast time-varying nature of the propagation channel, the uplink and downlink CSI are not identical. This issue can be addressed by deep learning (DL) based channel prediction techniques~\cite{Liu2022TCom, Wu2021JSAC}, and it is assumed that the BS has deployed an appropriate neural network to perfectly compensate the uplink–downlink CSI difference. During the design}, the transmitted symbols are carefully constructed so that the resulting interference contributes constructively to symbol detection. With such a design, the user receiver no longer needs to carry out conventional channel estimation or equalization procedures. Data symbols can be detected directly from the received signal, which significantly simplifies the receiver processing.

Traditional AFDM receivers often involve complicated signal processing procedures, resulting in considerable computational requirements. This can pose practical challenges for user devices with limited hardware resources. By transferring most of the processing tasks to the BS, the proposed framework effectively alleviates this burden on the user side and improves the practicality of AFDM systems. In addition, compared with the embedded pilot strategy adopted in the conventional AFDM scheme~\cite{Bemani2023TWC}, the proposed framework eliminates the need for frequent pilot insertion in the downlink transmission, thereby improving spectral efficiency.

\section{Downlink Probabilistic SLP-based Waveform Design}

In this section, we investigate the probabilistic SLP based waveform design for the downlink phase. We first introduce the principle of probabilistic SLP and formulate the corresponding waveform design problem. The resulting optimization problem is then reformulated into a more tractable form to be solved.

\subsection{Problem Formulation}
Due to multipath propagation and Doppler shifts, the received symbols may be displaced from their correct decision regions. By employing probabilistic SLP, the BS can pre-compensate the transmitted symbols such that the received symbols fall within the desired decision regions with high probability. In what follows, we design the downlink transmitted symbols in the AF domain based on probabilistic SLP technology. 

During the downlink transmission process, the symbol received on the $n$-th AF-domain subcarrier can be expressed as
\begin{equation}
	y_n = \mathbf{h}_n \mathbf{z} + \widetilde{{w}}_n,n=0,\cdots,N-1,
\end{equation}
where $\mathbf{h}_n$ denotes the $n$-th row of the channel matrix $\mathbf{H}_{\text{eff}}$, which is estimated at the BS during the first phase. \textcolor{black}{The vector $\mathbf{z}$ represents the designed signal obtained by applying probabilistic SLP to the original data symbol vector $\mathbf{x}$. Specifically, $\mathbf{z}=f\left(\mathbf{x}\right)$, where $f(\cdot)$ denotes the nonlinear precoding function that maps the intended symbols into a precoded signal adapted to the channel conditions.} The precoding objective is to ensure that each received symbol $y_n$ is correctly detected as the intended symbol $x_n$, drawn from a $4$PSK constellation set $\mathcal{X}$, i.e.,
\begin{equation}
	x_n \in \mathcal{X} = \left\{\text{exp}\left(\frac{j\pi\left(2i-1\right)}{4}\right),i=1,\dots,4\right\}.
\end{equation}

Due to the presence of AWGN, the received symbol $y_n$ follows a complex Gaussian distribution, i.e.,
\begin{equation}
	y_n \sim \mathcal{CN}(\mathbf{h}_n \mathbf{z},\sigma^2),
\end{equation}
where the real and imaginary components are independently distributed as
\begin{equation} \label{PDF of real}
	\text{Re}\{y_n\} \sim \mathcal{N}(\text{Re}\{\mathbf{h}_n \mathbf{z}\},\sigma^2/2),
\end{equation}
and
\begin{equation} \label{PDF of imag}
	\text{Im}\{y_n\} \sim \mathcal{N}(\text{Im}\{\mathbf{h}_n \mathbf{z}\},\sigma^2/2).
\end{equation}

Based on the above probabilistic characterization, the location of the received symbol can be analyzed from a statistical perspective. In the case of 4PSK modulation, the transmitted information is determined by the signs of the real and imaginary components of the received symbol. According to (\ref{PDF of real}), the probability that the real part of the received symbol is positive can be expressed as
\begin{align}
	\text{P}\left(\text{Re}\{y_n\}>0\right) & = \text{P}\left(\frac{\text{Re}\{y_n\}-\text{Re}\{\mathbf{h}_n \mathbf{z}\}}{\sigma/\sqrt{2}}>\frac{-\text{Re}\{\mathbf{h}_n \mathbf{z}\}}{\sigma/\sqrt{2}}\right) \notag \\
	& = 1 - \Phi\left(\frac{-\text{Re}\{\mathbf{h}_n \mathbf{z}\}}{\sigma/\sqrt{2}}\right) = \Phi\left(\frac{\text{Re}\{\mathbf{h}_n \mathbf{z}\}}{\sigma/\sqrt{2}}\right) \notag \\
	& = \frac{1}{2}\left[1+\text{erf}\left(\frac{\text{Re}\{\mathbf{h}_n \mathbf{z}\}}{\sigma}\right)\right],
\end{align}
where $\Phi\left(\cdot\right)$ denotes the cumulative distribution function (CDF) of the standard normal distribution and $\text{erf}\left(\cdot\right)$ denotes the Gauss error function.

Similarly, we have 
\begin{align}
	\text{P}\left(\text{Re}\{y_n\}<0\right) =\frac{1}{2}\left[1-\text{erf}\left(\frac{\text{Re}\{\mathbf{h}_n \mathbf{z}\}}{\sigma}\right)\right],
\end{align}
\begin{align}
	\text{P}\left(\text{Im}\{y_n\}>0\right) =\frac{1}{2}\left[1+\text{erf}\left(\frac{\text{Im}\{\mathbf{h}_n \mathbf{z}\}}{\sigma}\right)\right],
\end{align}
\begin{align}
	\text{P}\left(\text{Im}\{y_n\}<0\right) =\frac{1}{2}\left[1-\text{erf}\left(\frac{\text{Im}\{\mathbf{h}_n \mathbf{z}\}}{\sigma}\right)\right].
\end{align}
Since the real and imaginary parts are mutually independent, the probability that the received symbol $y_n$ lies in the first quadrant can be expressed as
\begin{align}
	\text{P}&\left(\text{Re}\{y_n\}>0,\text{Im}\{y_n\}>0\right)  \notag \\
	& = \text{P}\left(\text{Re}\{y_n\}>0\right)\text{P}\left(\text{Im}\{y_n\}>0\right) \notag \\
	& = \frac{1}{4}\left[1+\text{erf}\left(\frac{\text{Re}\{\mathbf{h}_n \mathbf{z}\}}{\sigma}\right)\right]\left[1+\text{erf}\left(\frac{\text{Im}\{\mathbf{h}_n \mathbf{z}\}}{\sigma}\right)\right].
\end{align}
The same reasoning applies to the other quadrants. Accordingly, the probability that the received symbol is correctly detected, denoted by $\Xi\left(\mathbf{h}_n \mathbf{z}\right)$, can be summarized as

\begin{align}
	\Xi(\mathbf{h}_n\mathbf{z})
	& =\frac{1}{4}\left[1+\sqrt{2}\operatorname{Re}\{x_n\}\operatorname{erf}\!\left(\frac{\operatorname{Re}\{\mathbf{h}_n\mathbf{z}\}}{\sigma}\right)\right] \notag\\
	&\left[1+\sqrt{2}\operatorname{Im}\{x_n\}\operatorname{erf}\!\left(\frac{\operatorname{Im}\{\mathbf{h}_n\mathbf{z}\}}{\sigma}\right)\right], x_n\in\left\{\frac{\pm1\pm j}{\sqrt{2}}\right\}.
	\label{RGD}
\end{align}

Next, we focus on the probability of symbol detection errors. For example, when the intended symbol $x_n$ lies in the first quadrant, the corresponding error probability can be expressed as
\begin{align}
	\text{P}_e&\left(\text{Re}\{y_n\}>0,\text{Im}\{y_n\}>0\right)  \notag \\ &=1-\text{P}\left(\text{Re}\{y_n\}>0,\text{Im}\{y_n\}>0\right) \notag \\
	& = 1- \frac{1}{4}\left[1+\text{erf}\left(\frac{\text{Re}\{\mathbf{h}_n \mathbf{z}\}}{\sigma}\right)\right]\left[1+\text{erf}\left(\frac{\text{Im}\{\mathbf{h}_n \mathbf{z}\}}{\sigma}\right)\right].
\end{align}
Therefore, for the intended transmitted symbol vector $\mathbf{s}$, the minimization of the bit error rate (BER) can be formulated as the following optimization problem
\begin{equation} \label{probp}
	\begin{aligned}
		\min_{\mathbf{z}} \quad & \max_{n} \{1-\Xi\left(\mathbf{h}_n \mathbf{z}\right)\} \\
		\text{s.t.} \quad &  \|\mathbf{z}\|_2^2 \leq P_m,
	\end{aligned}
\end{equation}
where $P_m$ denotes the transmit power budget.

\subsection{Problem Transformation and its Solution}
It is worth noting that the 
original problem in (\ref{probp}) can be reformulated as
\begin{equation} \label{probp1}
	\begin{aligned}
		\max_{\mathbf{z}} \quad & \min_{n}  T(\mathbf{h}_n \mathbf{z})\\
		\text{s.t.} \quad &  \|\mathbf{z}\|_2^2 \leq P_m, \\
		& T(\mathbf{h}_n \mathbf{z})=\left[1+\sqrt{2}\text{Re}\{x_n\}\operatorname{erf}\left(\frac{\operatorname{Re}\left\{\mathbf{h}_n \mathbf{z}\right\}}{\sigma}\right)\right]  \\
		& \left[1+\sqrt{2}\text{Im}\{x_n\}\operatorname{erf}\left(\frac{\operatorname{Im}\left\{\mathbf{h}_n \mathbf{z}\right\}}{\sigma}\right)\right],
	\end{aligned}
\end{equation}
where $T(\mathbf{h}_n \mathbf{z})$ characterizes the correct detection probability associated with the $n$-th symbol.

To facilitate subsequent optimization, we take the logarithm of $T(\mathbf{h}_n \mathbf{z})$, and the problem in (\ref{probp1}) can be equivalently transformed into
\begin{equation} \label{probp2}
	\begin{aligned}
		\max_{\mathbf{z}} \quad & \min_{n}  S(\mathbf{h}_n \mathbf{z})\\
		\text{s.t.} \quad &  \|\mathbf{z}\|_2^2 \leq P_m, \\
		& S(\mathbf{h}_n \mathbf{z})=\log\left[1+\operatorname{erf}\left(\frac{\text{Re}\{x_n\}\operatorname{Re}\left\{\mathbf{h}_n \mathbf{z}\right\}}{\sigma}\right)\right] \\
		& + \log\left[1+\operatorname{erf}\left(\frac{\text{Im}\{x_n\}\operatorname{Im}\left\{\mathbf{h}_n \mathbf{z}\right\}}{\sigma}\right)\right],
	\end{aligned}
\end{equation}
where $ S(\mathbf{h}_n \mathbf{z})=\log T(\mathbf{h}_n \mathbf{z})$. By exploiting the monotonicity of $\log(1+x)$ and $\operatorname{erf}(x)$, the objective can be approximated by focusing on the arguments of these functions. This leads to the following simplified problem
\begin{equation} \label{probp3}
	\begin{aligned}
		\max_{\mathbf{z}} \quad & \min_{n}  \{\text{Re}\{x_n\}\operatorname{Re}\left\{\mathbf{h}_n \mathbf{z}\right\}, \text{Im}\{x_n\}\operatorname{Im}\left\{\mathbf{h}_n \mathbf{z}\right\}\}\\
		\text{s.t.} \quad &  \|\mathbf{z}\|_2^2 \leq P_m,
	\end{aligned}
\end{equation}

By introducing an auxiliary variable $t$, the problem in (\ref{probp3}) can be further transformed into
\begin{equation} \label{probp4}
	\begin{aligned}
		\max_{\mathbf{z},t} \quad & t\\
		\text{s.t.} \quad &  \text{Re}\{x_n\}\operatorname{Re}\left\{\mathbf{h}_n \mathbf{z}\right\} \geq t, \\
		&  \text{Im}\{x_n\}\operatorname{Im}\left\{\mathbf{h}_n \mathbf{z}\right\} \geq t, \\
		& \|\mathbf{z}\|_2^2 \leq P_m.
	\end{aligned}
\end{equation}
The problem in (\ref{probp4}) is a SOCP, which can be efficiently solved using off-the-shelf convex optimization solvers such as CVX~\cite{cvx,gb08}. \textcolor{black}{As stated in~\cite{ben2001lectures}, the computational complexity of solving the SOCP in (\ref{probp4}) using an interior-point method is approximately $\mathcal{O}\left(N_{it}N^{3.5}\right)$ where $N_{it}$ denotes the iteration number. By solving its dual problem, the complexity can be reduced to $\mathcal{O}\left(N_{it}N^{2}\right)$~\cite{Li2018TWC}.}

\section{Numerical Results}

In this section, we evaluate the effectiveness of the proposed probabilistic SLP-based waveform design method in the downlink phase through Monte Carlo simulations. The simulation parameters are summarized in Table \ref{tab:sim_params}. The Doppler shifts for each path are randomly generated within the range $[-\alpha_{\text{max}}, \alpha_{\text{max}}]$. For AF-domain channel estimation, a Zadoff–Chu (ZC) sequence is employed as the pilot, and the channel is estimated using the algorithm proposed in~\cite{Tang2025ICCC}.

\begin{table}[t]
	\caption{Simulation Parameters}
	\label{tab:sim_params}
	\centering
	\begin{tabular}{|c|c|}
		\hline
		\textbf{Parameter} & \textbf{Value} \\
		\hline
		Subcarrier number & $N = 64$ \\
		\hline
		Carrier frequency & $f_c = 4~\text{GHz}$ \\
		\hline
		Subcarrier bandwidth & $\Delta f = 15~\text{kHz}$ \\
		\hline
		Total number of paths & $P = 3$ \\
		\hline
		Maximum normalized Doppler & $\alpha_{\max} = 1$ \\
		\hline
		Maximum normalized delay & $l_{\max} = 2$ \\
		\hline
		Channel coefficients & $h_i \sim \mathcal{CN}(0, 1/P)$ \\
		\hline
	\end{tabular}
\end{table}

\begin{figure}
	\centering
	\includegraphics[width=0.4\textwidth]{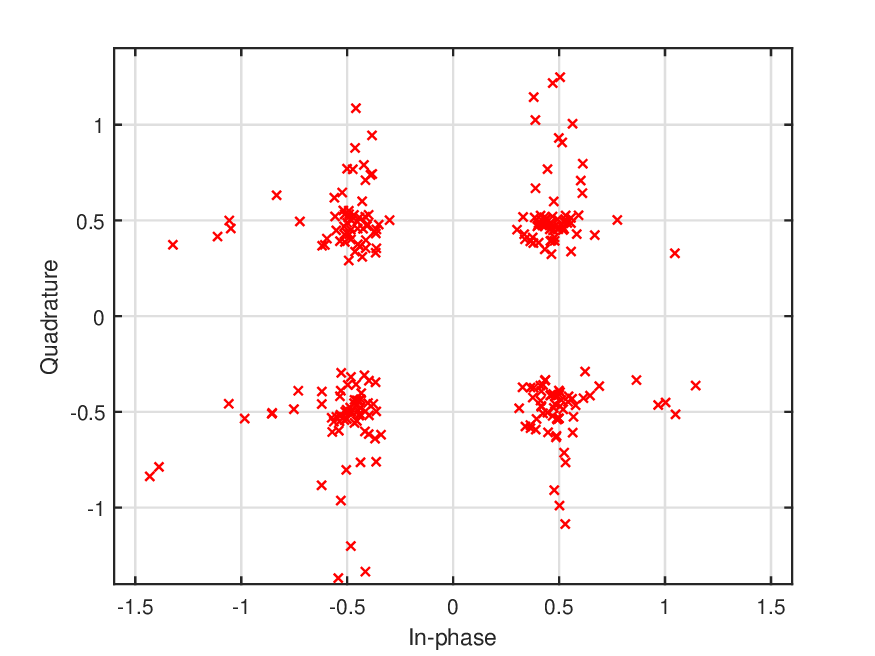}
	\caption{The constellation of the received symbol with probabilistic SLP design.}\label{fig:1}
\end{figure}

We first evaluate the performance of the proposed probabilistic SLP-based waveform design under the assumption of perfect CSI. The corresponding received symbol constellation at the user side is depicted in Fig.~\ref{fig:1}, where the signal to noise ratio (SNR) is set to 20 dB. By leveraging the perfect CSI, the proposed waveform scheme pre-adjusts the transmitted symbols to counteract channel-induced distortions, thereby increasing the likelihood that the received symbols fall into their correct decision regions. As observed, the received symbols are well concentrated within the four quadrants associated with the 4PSK constellation. This result demonstrates that the proposed waveform design effectively mitigates the impact of the channel, allowing reliable symbol detection at the user side with minimal processing and without the need for explicit channel estimation and equalization.

\begin{figure}
	\centering
	\includegraphics[width=0.4\textwidth]{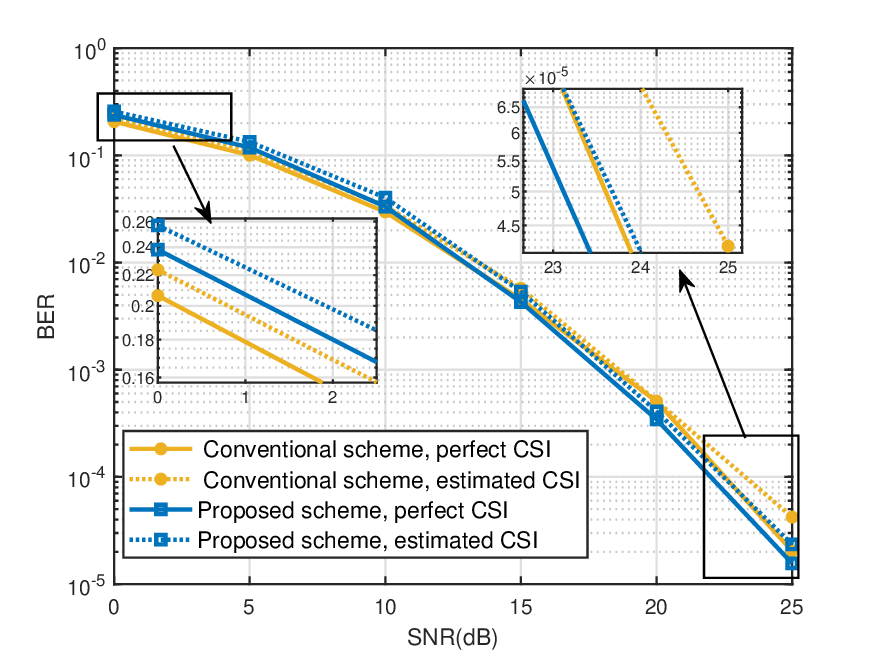}
	\caption{Downlink BER performance under different schemes with perfect CSI and estimated CSI.}\label{fig:2}
\end{figure}

As a step further, Fig. \ref{fig:2} illustrates the downlink BER performance of different schemes under both perfect CSI and estimated CSI. As observed, the proposed probabilistic SLP-based scheme achieves performance comparable to the conventional scheme across the entire SNR range, and exhibits a slight performance gain in the high-SNR regime. This improvement can be attributed to the fact that the proposed scheme exploits nonlinear optimization, thereby providing additional improvements compared to conventional linear equalization methods.
Comparing the perfect CSI and estimated CSI cases, it can be seen that the performance degradation caused by imperfect CSI is relatively minor for both schemes. \textcolor{black}{Notably, the proposed probabilistic SLP-based scheme maintains its performance advantage when using estimated CSI obtained from the highly accurate channel estimator represented in~\cite{Tang2025ICCC}, demonstrating both the scheme’s robustness to channel estimation errors and the effectiveness of the adopted estimator.} This indicates that the proposed framework remains effective in practical scenarios where perfect CSI is not available.
Overall, the results confirm that the proposed probabilistic SLP-based waveform design can achieve reliable performance while significantly reducing the receiver-side complexity, making it a promising solution for practical AFDM systems.

\begin{figure}
	\centering
	\includegraphics[width=0.4\textwidth]{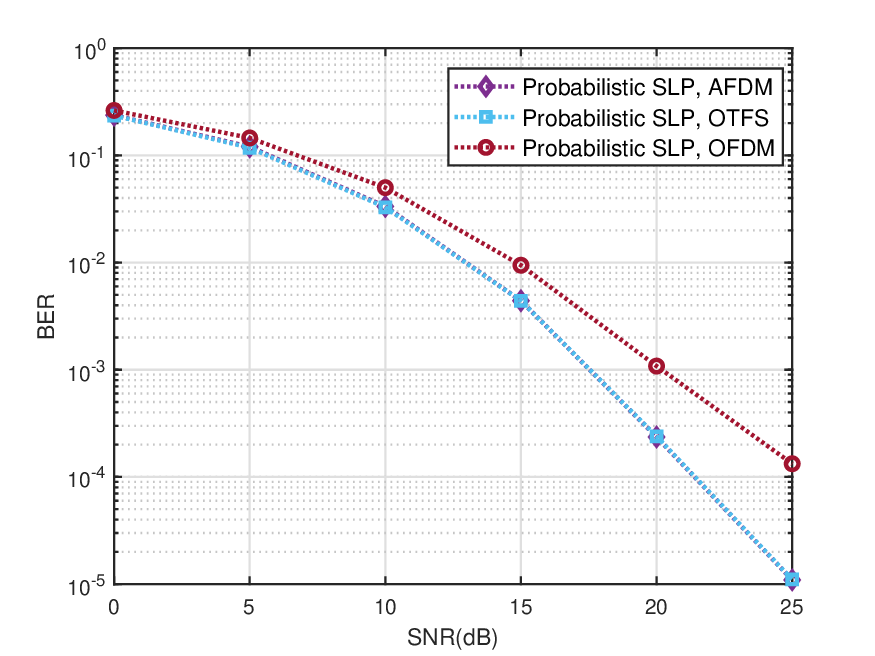}
	\caption{BER performance comparison among different modulation schemes.}\label{fig:3}
\end{figure}

\textcolor{black}{Finally, the proposed probabilistic SLP scheme is further generalized to both OFDM and orthogonal time frequency space (OTFS) frameworks to verify its scalability, as illustrated in Fig.~\ref{fig:3}. The results indicate that the AFDM- and OTFS-based implementations exhibit almost indistinguishable performance, and both consistently surpass the OFDM-based scheme. This performance gain can be attributed to the ability of AFDM and OTFS to effectively harness the inherent diversity in doubly selective channels. Overall, these observations confirm that the proposed approach maintains strong adaptability across different waveform architectures.}

\section{Conclusion}

In this paper, we proposed a probabilistic SLP-based waveform design for AFDM systems to alleviate the receiver-side computational burden. By exploiting the uplink CSI, the BS designs the transmit symbols to minimize the probability of erroneous detection, enabling direct symbol detection without channel estimation or equalization at the receiver. Simulation results show that the proposed scheme achieves BER performance comparable to conventional methods under both perfect and estimated CSI, while significantly reducing receiver complexity. These results demonstrate the effectiveness and practicality of the proposed framework for AFDM system.

\section*{Acknowledgement}
This work was supported by the National Natural Science Foundation of China (NSFC) under Grant 62301032 and 62471039. The work of Shuangyang Li was supported in part by the European Research Council (ERC) under the ERC Starting Grant No. 101220383 (Foundations of Delay Doppler Communications and Sensing, FUNDOCS).

\bibliographystyle{IEEEtran}%
\bibliography{bib/bibfile}
\end{document}